\documentclass[%
 aps,
 pra,%
 amsmath,amssymb,
reprint,%
superscriptaddress,
]{revtex4-1}

\usepackage{graphicx}
\usepackage{bm}
\usepackage{dcolumn}
\usepackage[utf8]{inputenc}
\usepackage{textcomp}
\usepackage{comment}

\usepackage{booktabs}
\usepackage{rotating}
\usepackage{multirow}
\usepackage{subfigure}
\usepackage[normalem]{ulem}
\usepackage{amsmath}
\usepackage[usenames,dvipsnames,svgnames,table]{xcolor}
\usepackage{booktabs}
\usepackage{rotating}
\usepackage{multirow}
\usepackage{subfigure}
\usepackage{adjustbox}
\usepackage{graphicx}
\usepackage{dcolumn}
\usepackage{bm}
\usepackage{lineno}
\usepackage{amsmath}
\usepackage{url}

\usepackage{tabularx}

\usepackage{color}
\usepackage{hyperref}
\hypersetup{
     colorlinks   = true,
     citecolor    = blue,
     linkcolor    = blue,
     urlcolor     = blue
}

\usepackage{url,hyperref}

\usepackage{color}
\usepackage[usenames,dvipsnames,svgnames,table]{xcolor}

\newcommand{\FD}[1]{\textcolor{Blue}{#1}}

\usepackage{amssymb}
\usepackage{amsmath}
\begin{document}
\frenchspacing



\title{Polarization characteristics of adatoms self-diffusing on metal surfaces under high electric fields}

\author{Ekaterina Baibuz}
\affiliation{Helsinki Institute of Physics and Department of Physics, P.O. Box 43 (Pietari Kalmin Katu 2), FI-00014 University of Helsinki, Finland}
\author{Andreas Kyritsakis}
\affiliation{Institute of Technology, University of Tartu, Nooruse 1, 50411 Tartu, Estonia}
\email{andreas.kyritsakis@ut.ee}
\author{Ville Jansson}
\affiliation{Helsinki Institute of Physics and Department of Physics, P.O. Box 43 (Pietari Kalmin Katu 2), FI-00014 University of Helsinki, Finland}
\author{Flyura Djurabekova}
\affiliation{Helsinki Institute of Physics and Department of Physics, P.O. Box 43 (Pietari Kalmin Katu 2), FI-00014 University of Helsinki, Finland}

\begin{abstract}
Although atomic diffusion on metal surfaces under high electric fields has been studied theoretically and experimentally since the 1970s, its accurate and quantitative theoretical description remains a significant challenge. In our previous work, we developed a theoretical framework that describes the atomic dynamics on metal surfaces in the presence of an electric field in terms of the local polarization characteristics of the surface at the vicinity of a moving atom. 
Here, we give a deeper analysis of the physics underlying this framework, introducing and rigorously defining the concept of the effective polarization characteristics (permanent dipole moment $\mu$ and polarizability $\alpha$) of a moving atom on a metal surface, which are shown to be the relevant atomic quantities determining the dynamics of a moving atom via a compact equation. 
We use density functional theory (DFT) to calculate $\mu$ and $\alpha$ of a W adatom moving on a W \{110\} surface, where additional adatoms are present in its vicinity. We analyze the dependence of $\mu$ and $\alpha$ and hence the migration barriers under electric fields on the local atomic environments (LAE) of an adatom. 
We find that the LAE significantly affects $\mu$ and $\alpha$ of a moving atom in the limited cases we studied, which implies that further systematic DFT calculations are needed to fully parameterize surface diffusion in terms of energy barriers for long-term large scale simulations, such as our recently developed Kinetic Monte Carlo model for surface diffusion under electric field.
\end{abstract}
\maketitle

\section{Introduction}

When a high electric field is applied on a metal surface, it modifies significantly the dynamic behavior and especially the diffusion characteristics of its atoms, leading to significant surface modifications
\cite{tsong1972measurements,tsong1971measurement,tsong1975direct,kellogg1993electric,feibelman2001surface,fujita2007mech, kyritsakis2018thermal, Parviainen2011, vigonski2015molecular}. These effects can be exploited to manipulate the surface atoms at an
atomic level \cite{stroscio_atomic_1991, whitman_manipulation_1991} by causing biased diffusion due to electric field effects.
Nanostructures such as islands of adatoms and surface mounds, can
be formed under high electric fields \cite{stroscio_atomic_1991}. 
A similar method that exploits these field effects has been proposed for fabricating atomic switches for the next generation beyond-von-Neumann computers \cite{hasegawa_atomic_2012}.
High fields are also used in combination with heating to sharpen the
metal tips \cite{yanagisawa2009optical, yanagisawa2016laser} to be used in various applications such as electron, ion and field emission microscopy \cite{tsong_mechanisms_2005,morgan_notte_hill_ward_2006,GarciaCoherent1989,JensenImage, gomer1992field, harp1990atomic, MEVVA1985, lai2017xenon,Muller1965field, muller1956field},
while in Atom Probe Tomography (APT) \cite{miller_atom-probe_2014, miller2012atom, Perea2009direct, Kelly2007invited}, the high electric field
removes atoms from a metal surface through field evaporation, thus
producing an image of the material structure.
On the other hand, our recent theoretical results \cite{jansson2020growth} indicate that high electric fields may cause the growth of surface nanotips that may be responsible for vacuum breakdowns (also called vacuum arcs), which are detrimental to the function of many devices, spanning from nano- and microelectronics \cite{zhou2019direct,de2006multipactor, rozario1994investigation, lyon2013gap, ducharme2009inside, sterling2013increased} to  vacuum interrupters \cite{slade2018vacuum} and
X-ray tubes \cite{latham1995high}, up to large-scale apparatuses such as fusion
reactors \cite{MCCRACKEN19803} and existing and future particle accelerators \cite{Descoeudres2009, antoine2012electromigration, clic2016, engelberg2018stochastic}

\begin{figure}[!ht]
  \caption{Charge redistribution is induced by a positive 1 GV/m applied electric field (anode, on the right) and a negative 1 GV/m field (cathode, on the left). Redistribution was obtained as a difference between the charge density of the system under electric field and the charge density of the system without electric field as calculated with DFT. Atomic positions of the slab, which was relaxed under the field first, were fixed and only the electronic relaxation was performed for the system without a field for illustration purposes. The open surface of the slab is \{110\} oriented. Cyan and magenta-colored areas correspond to increased and decreased electron densities, respectively,
that exceed 0.1\% of the maximum electron density of the reference system without field and the adatom}
  \centering
    \includegraphics[width=0.48\textwidth]{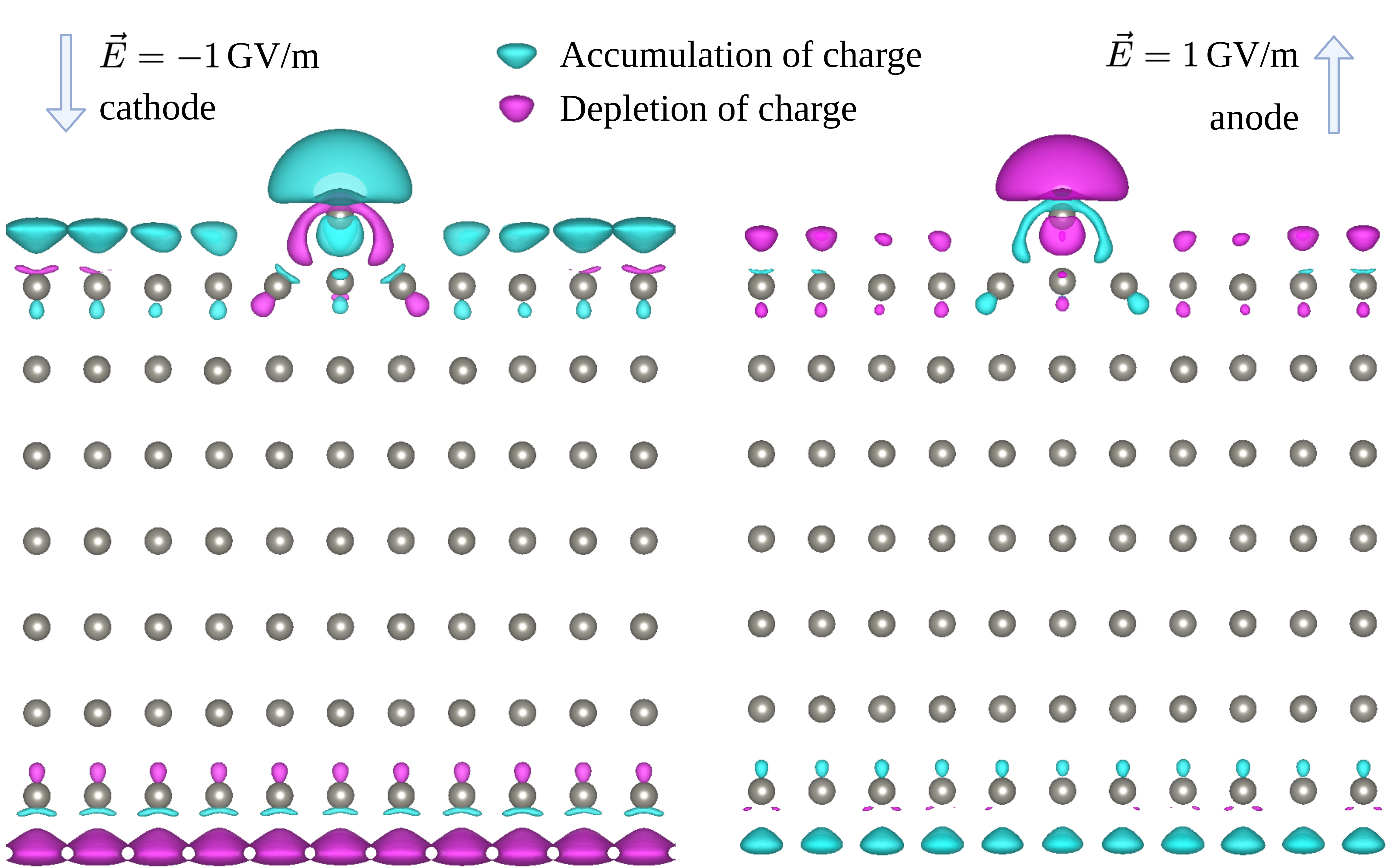}
\label{fig:anode+cathode}
\end{figure} 

Since the 1960s, the modification of the surface dynamics by the presence of a high electric field has been theoretically described in terms of the polarization characteristics (permanent dipole moment and polarizability) of the surface atoms. The field evaporation theory works \cite{drechsler1957kristallstufen,muller1960field,tsong1969effects,tsong1970fieldevaporation,tsong1970fieldadsorption,tsong1968mechanism,brandon1965structure,gomer1963theoryoffield} introduced the concept of \textit{effective adatom polarizability} into the binding energy (removal work) equation as the coefficient of a square-dependence term on the applied electric field. However, the term effective polarizability was not rigorously defined in these works.

Similar concepts were introduced by Tsong et al. \cite{tsong1972measurements, tsong1971measurement,tsong1975direct}, to analyze the lowering of the energy barrier for surface diffusion of a W adatom on a W \{110\} surface. 
In their experiments, it was observed that the diffusion of the adatom is biased towards the edges of a W\{110\} surface, where the local electric field is higher due to the sharp edges. 
The observed dependence of the biased diffusion characteristics on the applied electric field led to the development of a theory that describes how the migration barriers are affected by a non-uniform applied electric field, based on similar concepts used in the field evaporation theory.
Tsong and Kellogg (TK) \cite{tsong1975direct} wrote the energy of an adatom $E$ under high electric field $F$ as

\begin{equation} \label{eq:tsong_energy}
    E(F) = E(F=0) - \mu F - \frac{1}{2} \alpha F^2 \textrm{.}
\end{equation}

The parameters $\mu$ and $\alpha$ were called the \textit{surface-induced permanent dipole moment} and \textit{polarizability of an adatom} correspondingly, defining the dipole moment of an adatom under field as $p = \mu + \alpha F$.

TK  suggested that the origin of the polarizability of an adatom is in the charge transfer from surface atoms to the adatom.
Although they clearly stated that $\mu$ and $\alpha$ ``are necessarily different from those for a free atom'' \cite{tsong1975direct} and depend on the adatom's position on the surface, their analysis essentially treats the migrating adatom as if it were isolated, ignoring the charge redistribution induced in its vicinity.

In our recent work \cite{kyritsakis2019atomistic}, we showed that the modification of the migration barriers ---  and in general the potential energy landscape under a field ---  is driven by a charge redistribution not only on the atom but rather the whole system that includes the atoms in the vicinity of the migrating one (see Fig.\ref{fig:anode+cathode} for charge redistribution).
Therefore, we had to consider the modification of the energy of the whole system that includes the migrating atom and its local atomic environment (LAE), which is

\begin{equation} \label{eq:Free}
  E(F) = E(F=0) - \mathcal{M} F - \frac{1}{2} \mathcal{A} F^2 \textrm{.}
\end{equation}

Here, $\mathcal{M}$ and $\mathcal{A}$ are the \textit{systemic} permanent dipole moment and polarizability, respectively. We use the calligraphic letters to distinguish between the systemic and other entities. All entities in Eq. \eqref{eq:Free} are scalars. From now on we will omit the word ``permanent'', always meaning the dipole moment without electric field unless specifically stated otherwise. Systemic polarization characteristics are determined by the charge redistribution of the whole surface --- adatom and surface atoms included.
$\mathcal{M}$ and $\mathcal{A}$ for an atomic system such as the one in Fig. \ref{fig:anode+cathode} can be calculated directly using Density Functional Theory (DFT), according to the following linear dependency: $\mathcal{P} = \mathcal{M} + \mathcal{A}F$, where  $\mathcal{P}$ is the systemic dipole moment under the electric field $F$. 

Equation \eqref{eq:Free} and the other results from Ref. \cite{kyritsakis2019atomistic} were implemented into a Kinetic Monte Carlo (KMC) model which we used to study how the coupling between an applied electric field and the atom migration barriers would affect the surface evolution of a W surface \cite{jansson2020growth}.
Although the systemic parameters $\mathcal{M}$ and $\mathcal{A}$ can be assumed to change after every atomic jump in the system, the model approximated all $\mathcal{M}$ and $\mathcal{A}$ to be constant, using for all jumps the DFT-calculated values for a single W adatom jump on a W\{110\} surface. The effects of the LAE were thus ignored. Nevertheless, the model was able to predict the growth of surface nanotips due to biased diffusion at sufficiently high electric fields. The nanotip-growth was observed even if the $\mathcal{M}$ and $\mathcal{A}$ parameter values were varied with $\pm$20 \%. The model also qualitatively reproduced the W surface faceting patterns observed under electric fields in experiments by Fujita et al. \cite{fujita2007mech}.
Nevertheless, it remains an open question to what degree the LAE of a migrating atom affects its effective polarization characteristics and subsequently the diffusion. 

In this article, we connect the concepts of atomic dipole moment and polarizability used in the TK's theory \cite{tsong1975direct} (Eq. \eqref{eq:tsong_energy}) with the systemic parameters introduced in our previous work \cite{kyritsakis2019atomistic} (Eq. \eqref{eq:Free}), by introducing and rigorously defining the \textit{effective dipole moment} and \textit{effective polarizability} (the effective polarization characteristics) of the adatom, which are the relevant atomic quantities that determine the dynamics of a moving atom on a metal surface under high electric field.
We first approximate these quantities with the partial charges, treating the charge distribution of an adatom as if it was isolated from its neighbors. 
We justify the effect of the vicinity of the adatom on its effective polarization characteristics and prove that these cannot be approximated with partial polarization. 
Then, keeping in mind the parameterization of KMC simulations for the surface evolution under electric field via diffusion, we explore how the LAE of a migrating atom changes its effective polarization characteristics and affects the energy barriers under an electric field.

\section{Theory}

\subsection{Effective atomic polarization characteristics}\label{subsec:theory_effective}
Let us begin with defining the concept of \textit{effective atomic} polarization characteristics as the difference in \textit{systemic} polarization characteristics when introducing an adatom. In this definition, the effective dipole moment $\mu_x$ and the effective polarizability $\alpha_x$ of the adatom at an arbitrary position $x$ on the surface can be calculated as follows:

\begin{equation} \label{eq:effective_characteristic}
\begin{split}
\mu_x \equiv  \mathcal{M_{\text{x}}} - \mathcal{M_{\text{r}}}, 
\\
\alpha_x \equiv \mathcal{A_{\text{x}}} - \mathcal{A_{\text{r}}} , 
\end{split}
\end{equation}

where $\mathcal{M_{\text{x}}}$ and $\mathcal{A_{\text{x}}}$ are systemic dipole moment and polarizability of the surface with an adatom at the position $x$;  $\mathcal{M_{\text{r}}}$ and $\mathcal{A_{\text{r}}}$ are the systemic polarization characteristics of the \textit{reference} system, i.e the same surface without the adatom.

Thus, the effective atomic dipole moment $\mu_x$ reflects the charge redistribution in the system due to the presence of an adatom, and the effective atomic polarizability $\alpha_x$ defines how this charge redistribution is affected by the electric field. 

Keeping in mind that the effective quantities introduced above in Eq. \eqref{eq:Free}, the migration barrier of an adatom under an electric field  can be written as

\begin{equation} \label{eq:barrier_effective}
    E_m = E_m(0) - \Delta \mu F_l - \frac{\Delta \alpha}{2}F_l^2 - \mu_s \Delta F - \alpha_s F_l \Delta F \textrm{,}
\end{equation}

where $\Delta \mu = \mu_s - \mu_l$, $\Delta \alpha = \alpha_s - \alpha_l$, $\mu_s$  is the effective dipole moment of the adatom at the saddle position (denoted $\mathcal{M_{\text{sr}}}$ in \cite{kyritsakis2019atomistic}), $\mu_l$ is the effective dipole moment of the adatom at the lattice site (denoted $\mathcal{M_{\text{lr}}}$ in \cite{kyritsakis2019atomistic}), 
$\alpha_s$ is the effective polarizability of the adatom at the saddle point (denoted $\mathcal{A_{\text{sr}}}$ in \cite{kyritsakis2019atomistic}),
$\alpha_l$  is the effective polarizability at the lattice site (denoted $\mathcal{A_{\text{lr}}}$ in \cite{kyritsakis2019atomistic}).

Equation \eqref{eq:barrier_effective} is to its format similar to the equation (9) of Ref. \cite{tsong1975direct} by TK, but the dipole moment and polarizability of the moving adatom are defined differently. Since the adatom is  part of the surface and cannot be considered independently in sense of its electronic properties, in Eq. \eqref{eq:barrier_effective} we used the relative ("effective") dipole characteristics, rigorously defined in terms of the systemic quantities introduced in [51].

Equation \eqref{eq:barrier_effective} can be generalized to obtain the potential energy change of any (small) movement of an atom on a metal surface under a field by substituting the saddle and lattice sites by any initial and final points, correspondingly.
The above definition of the effective atomic polarization characteristics is general and does not depend on the position of the atom.
Therefore, the potential energy $U$ of any atom at any point near the surface can be written as

\begin{equation} \label{eq:potential}
    U = U_0 - \mu_x F - \frac{1}{2} \alpha_x F^2
\end{equation}

and the force exerted on the atom as

\begin{equation} \label{eq:force}
    -\nabla U = -\nabla U_0 + F \nabla \mu_x + \frac{F^2}{2} \nabla \alpha_x + \mu_x \nabla F + \alpha_x F \nabla F \textrm{,}
\end{equation}

where $U_0$ is the atomic potential energy ($F=0$), and $\mu_x, \alpha_x$ are the position-dependent effective atomic dipole moment and polarizability of the moving atom.

We have now shown that the dynamics of atoms on metal surfaces under a high electric field can be described with a compact formula in terms of the rigorously defined effective atomic polarization characteristics. In the following sections, we will calculate $\mu_x$ and $\alpha_x$ for various atomic configurations (LAE) with DFT calculations and analyze their behavior.
However, before we enter this analysis, we shall explore the concept of partial atomic polarization characteristics and its relation to the concept of systemic $\mathcal{M}$ and $\mathcal{A}$, and thus effective $\mu_x$ and $\alpha_x$.

\subsection{Partial atomic dipole moment}
As effective polarization characteristics are represented by the differences between the system with the adatom at a certain position and a reference system where the adatom is absent, another approach of calculating $\mu_x$ and $\alpha_x$ could be to treat the adatom as if it were isolated as suggested by TK in their research \cite{tsong1972measurements, tsong1971measurement,tsong1975direct}. This approach takes into account the charge redistribution on the adatom only, thus  approximating the effective polarizabilities and the effective dipole moments in Eq. \eqref{eq:barrier_effective} with the dipole moments and the polarizabilities of the \textit{partial} charge density of the adatom. We will call the dipole moment and polarizability calculated from the partial charge density as the \textit{partial dipole moment} and \textit{partial polarizability}, respectively.  

\begin{figure}[h!]
  \caption{Partial dipole moments a) and partial charges b) of the adatom and its neighbors as obtained with DFT calculations and Bader analysis.}
  \centering
    \includegraphics[width=0.45\textwidth]{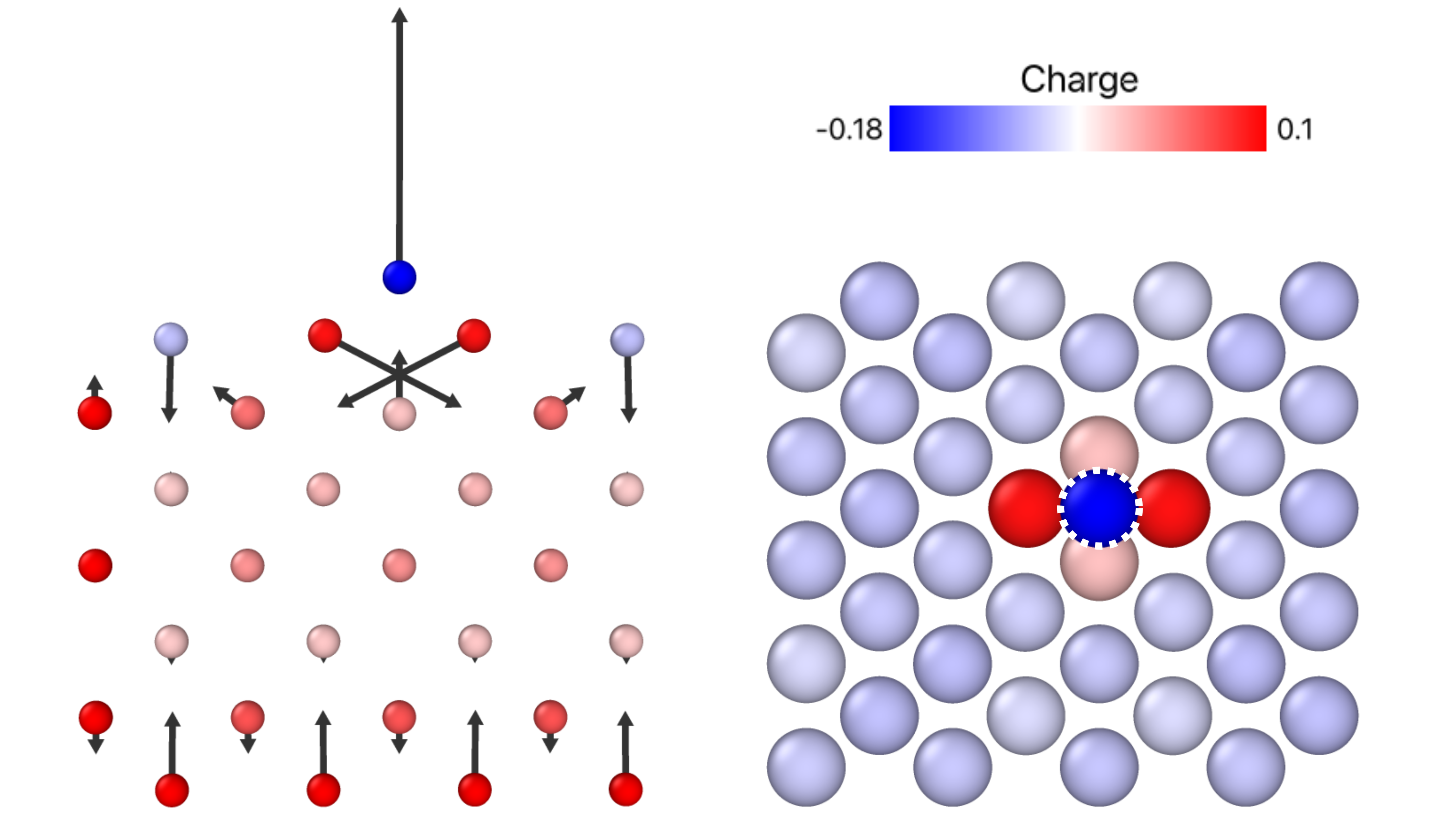}
\label{fig:charges_dipoles}
\end{figure}

We define the partial dipole moment of the adatom with the following formula:

\begin{equation}
  p^{(a)} = \int_{G^{(a)}} \rho (\vec{r})(\vec{r}-\vec{r_0})d\vec{r},
\label{eq:partial_mu}
\end{equation}

where the volume of integration $G^{(a)}$ corresponds to the partial charge density of the adatom $(a)$, i.e. the charge density in the immediate vicinity of the adatom; $r_0$ is the center coordinates of the adatom.

The systemic dipole moment of a system of atoms (under arbitrary field conditions) can be expressed in terms of partial charges assigned to each atom as

\begin{equation}
  \mathcal{\vec{P}} = \sum_i \vec{p^{(i)}} + \sum_i q^{(i)} \vec{R_i} \textrm{,}
  \label{eq:dipole_sum}
\end{equation}
where index $i$ runs over all atoms, $\vec{p^{(i)}}$ is the partial dipole moment and $q^{(i)}$ the partial charge on the $i$-th atom, which is located at the position $\vec{R_i}$ from the center of mass of the system.

Figure \ref{fig:charges_dipoles} illustrates Eq. \eqref{eq:dipole_sum} by showing partial dipole moment vectors and partial charges of individual atoms in the slab with one adatom system.

The approach to approximate the effective polarization characteristics with the partial ones would significantly lower the amount of needed calculations for obtaining $\mu_x$ and $\alpha_x$ comparing to the approach presented in the previous section, where four systemic values ($\mathcal{M_{\text{x}}}$, $\mathcal{M_{\text{r}}}$, $\mathcal{A_{\text{x}}}$, $\mathcal{A_{\text{r}}}$ ) are needed to calculate the effective polarization characteristics. Although in Sec. \ref{subsec:results_partial} we show that partial polarization characteristics are insufficient to describe the diffusion under electric field, we believe it adds up to the discussion on how the vicinity (and LAE) of the adatom modifies the electric field effect on its migration barriers.

\section{Methods}
\subsection{Systems under the study}
\begin{figure}[!h]
  \caption{ Studied systems of W adatoms on a W \{110\} surface: \textit{1nn} is a slab with two adatoms at lattice sites within the first-nearest neighbor (nn) distance from each other; \textit{2nn} - within the second-nn distance; \textit{3nn} - within the third-nn distance; \textit{4nn} - within the fourth-nn distance; \textit{3 adatoms} is a slab with three adatoms clustered together; \textit{4 adatoms} - a slab with a four adatoms island. The color of each atom corresponds to its charge, which was extracted with the Bader analysis from the charge density outputted by VASP. The color scale is the same as in Fig. \ref{fig:charges_dipoles}. Only adatoms and surface layer atoms are shown. Adatoms are highlighted with dashed borders. Unit cells of VASP calculations are shown with rectangles.}
  \centering
    \includegraphics[width=0.48\textwidth]{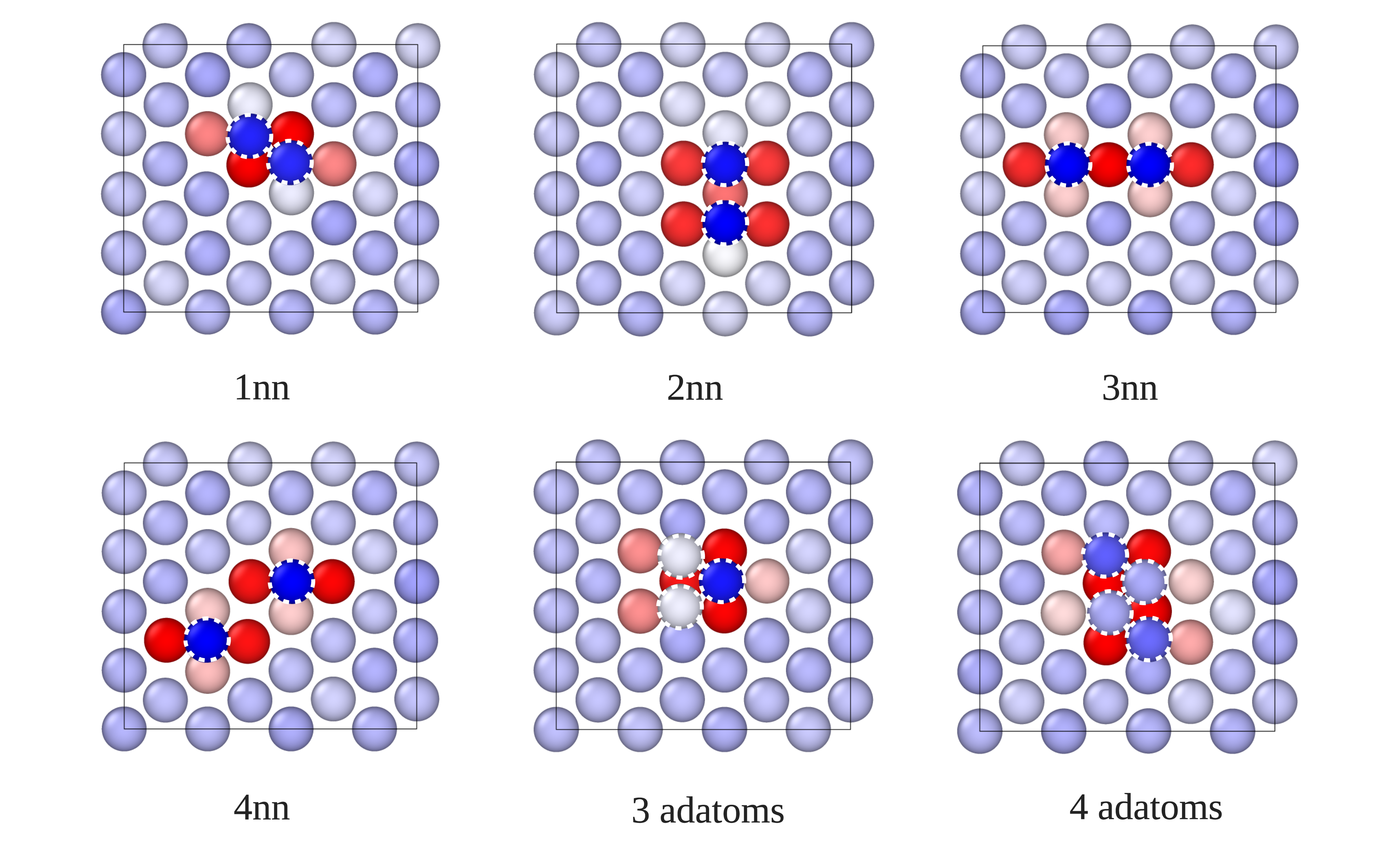}
\label{fig:systems}
\end{figure}
In this work, we utilized the results of the DFT simulations of a W adatom on a W \{110\} surface published in \cite{kyritsakis2019atomistic}. 
This system of a slab with an adatom on top is denoted as the \textit{adatom} system in this work. 
The \textit{adatom} system as it appears in our DFT simulations is depicted in Fig. \ref{fig:anode+cathode} and \ref{fig:charges_dipoles}.

We run the DFT calculations of various local atomic environments of adatoms on the W \{110\} surface. Fig. \ref{fig:systems} shows the systems we considered in this work and their notations. We studied how the effective polarization characteristics of an adatom change in the presence of other adatoms.
 In particular, we studied the effect of the nearest neighbors (nn) located at the first-nn (\textit{1nn} in Fig. \ref{fig:systems}), second-nn (\textit{2nn} in Fig. \ref{fig:systems}), third-nn (\textit{3nn} in Fig. \ref{fig:systems}), and fourth-nn (\textit{4nn} in Fig. \ref{fig:systems}) lattice positions from the adatom. We also studied clusters of three and four adatoms (\textit{3 adatoms} and \textit{4 adatoms} systems in Fig. \ref{fig:systems}).

\subsection{Partial atomic dipole moment and polarizability calculations} \label{sec:methods_partial}
To calculate partial polarization characteristics, we took the charge distribution of the \textit{adatom} system as calculated by DFT in \cite{kyritsakis2019atomistic} and separated the atomic charges, i.e. partitioned the global charge distribution around each atom. For charge partitioning, we used the Bader analysis \cite{tang2009grid,sanville2007improved,yu2011accurate} method as implemented by Henkelman et al. \cite{henkelman2006fast}.

\begin{figure}[t]
  \caption{Partial polarization characteristics calculations. On the left: partial charge density around the adatom extracted with the Bader analysis from the charge density of the whole system obtained with DFT. On the right: partial dipole moment $p^{(a)}$ of the adatom at lattice and saddle point positions (marked with dots) under the applied electric field $F$ as calculated by DFT, Bader analysis and Eq. \eqref{eq:partial_mu}. Dashed line is a linear fit to $p^{(a)} = m^{(a)} + a^{(a)} F$}
  \centering
    \includegraphics[width=0.48\textwidth]{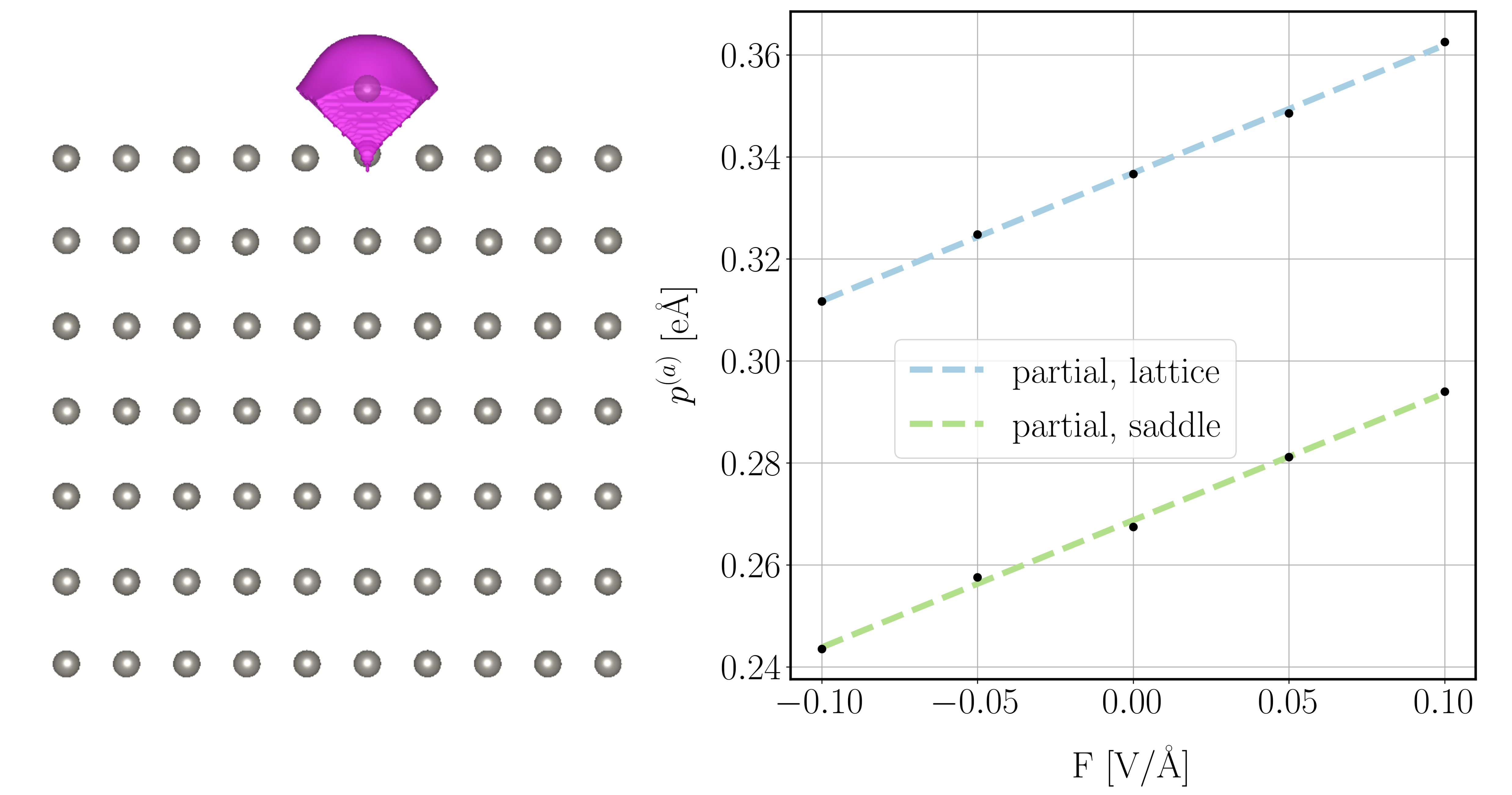}
\label{fig:bader}
\end{figure}

We calculated partial dipole moments of both saddle and lattice positions of the adatom for several values of the electric field. The partial dipole moments under electric field were calculated using Eq. \eqref{eq:partial_mu}. The partial charge densities of the adatom $G^{(a)}$ at saddle and lattice positions were determined by the Bader analysis for each case of the electric field.  
The partial dipole moment $m^{(a)}$ and partial polarizability $a^{(a)}$ of the adatom were then extracted with a linear fit to the equation $p^{(a)} = m^{(a)} + a^{(a)} F$, where $p^{(a)}$ is the partial dipole moment under a field $F$ as calculated by Eq. \eqref{eq:partial_mu}.

We then repeated these calculations for each atom in the \textit{adatom} system to obtain a set of vectors of partial dipole moments $p^{(i)}$ for Fig. \ref{fig:charges_dipoles}. Partial charge values $q^{(i)}$ for each atom were outputted by the Bader analysis software. 
In the same manner, the partial charges were obtained for all the systems shown in Fig. \ref{fig:systems}.

Fig. \ref{fig:bader} illustrates the partial polarization characteristics calculations. On the left is an example of the  partial charge density volume of the adatom, which was extracted from the charge density of the whole system calculated with DFT. Such a partitioned volume around the adatom was obtained for every field value to calculate $p^{(a)}$ of the adatom at lattice and saddle point positions (marked with dots on the right graph). The graph on the right shows the linear dependence of the $p^{(a)}$ vs. the  applied electric field $F$, from which $m^{(a)}$ and $a^{(a)}$ values can be obtained.

\subsection{Density Functional Theory calculations of systemic dipole moments and polarizabilities} 
To extract the polarization characteristics of the systems in Fig. \ref{fig:systems}, we performed DFT calculations of a slab with W adatoms on top of W \{110\} surface with and without an applied electric field. 

DFT calculations were performed with the Vienna ab initio simulation package (VASP) and its corresponding ultrasoft-pseudopotential database \cite{kresse1993ab,kresse1996efficiency,kresse1996efficient,vanderbilt1990soft,pasquarello1992ab,laasonen1993car,kresse1994norm}. 
For consistency, we used the same parameters and schemes for the DFT calculations as in \cite{kyritsakis2019atomistic}, which resulted  in a 1 meV convergence. 
In the calculations, we used the Perdew-Burke-Ernzerhof \cite{perdew1996generalized} generalized gradient approximation (GGA) functional, Blocked Davidson iteration scheme \cite{davidson1983methods}, the Methfessel-Paxton smearing scheme \cite{methfessel1989high} and the Kerker mixing scheme \cite{kerker1981efficient}. The cut-off energy of 600 eV was set for the plane wave basis. 
VASP uses the periodic slab model for surface simulations. For all slab systems under study, we used 8 monolayers of atoms in the $x$ direction, 10 in
the $y$, and 7 monolayers in the $z$ direction with an additional layer of adatoms and 24 $\textrm{\AA}$ of vacuum on top. The vacuum height is measured from the highest fully occupied atomic layer. Ionic relaxation with the conjugate-gradient method (see e.g. \cite{press1996numerical}) was used for simulations of systems in Fig. \ref{fig:systems}. A Gamma-centered k-grid was used for all the calculations. $7 \times 7 \times 1$ k-grid was used in all cases. The electric field is implemented in VASP through the artificial dipole sheet in the middle of the vacuum according to the scheme proposed in \cite{neugebauer1992adsorbate}. Dipole corrections were also used in the case of zero-field calculations to counteract the dipole interactions between repeated slabs (using the LDIPOL = .True. keyword in VASP). The relaxation was stopped when energies converged within 0.01 meV

The systemic polarizability and dipole moment values for each system in Fig. \ref{fig:systems} were calculated from the ground state energy vs applied field curves, $E$--$F$, which were fitted using Eq. \eqref{eq:Free}. 
To reach sufficient accuracy for polarizability, DFT calculations were run for five to seven values of the applied electric field up to 1.5 GV/m, both in the anode and the cathode directions (outwards and inwards the surface, respectively). 

Table \ref{table:systemic} lists the obtained values of systemic polarizabilities and dipole moments along with the error estimates. The error estimate of each value corresponds to the standard deviation, which was obtained from the least square  fit covariance matrix.

\subsection{Saddle point calculations}\label{subsec:methods_saddle} 

\begin{figure}[!t]
  \caption{Adatom jumps to the first nearest neighbor lattice positions in various LAE. Surface atoms are shown in yellow color, adatoms --  red. Dashed lines indicate the planes along the jumps. Forward jumps  "$\rightarrow$" are from the initial configuration on the right to the final one on the left. Reverse jumps $\leftarrow$ are in the opposite direction. The initial state of \textit{3 adatoms, forward} jump (on the right) corresponds to the \textit{3 adatoms} system in Fig. \ref{fig:systems}; initial state of \textit{3 adatoms, reverse} jump (on the left) is a system after one of the adatoms from \textit{3 adatoms} system has jumped away from the other two. \textit{3 adatoms, reverse} is then the jump of the apart adatom towards the other two.}
  \centering
    \includegraphics[width=0.3\textwidth]{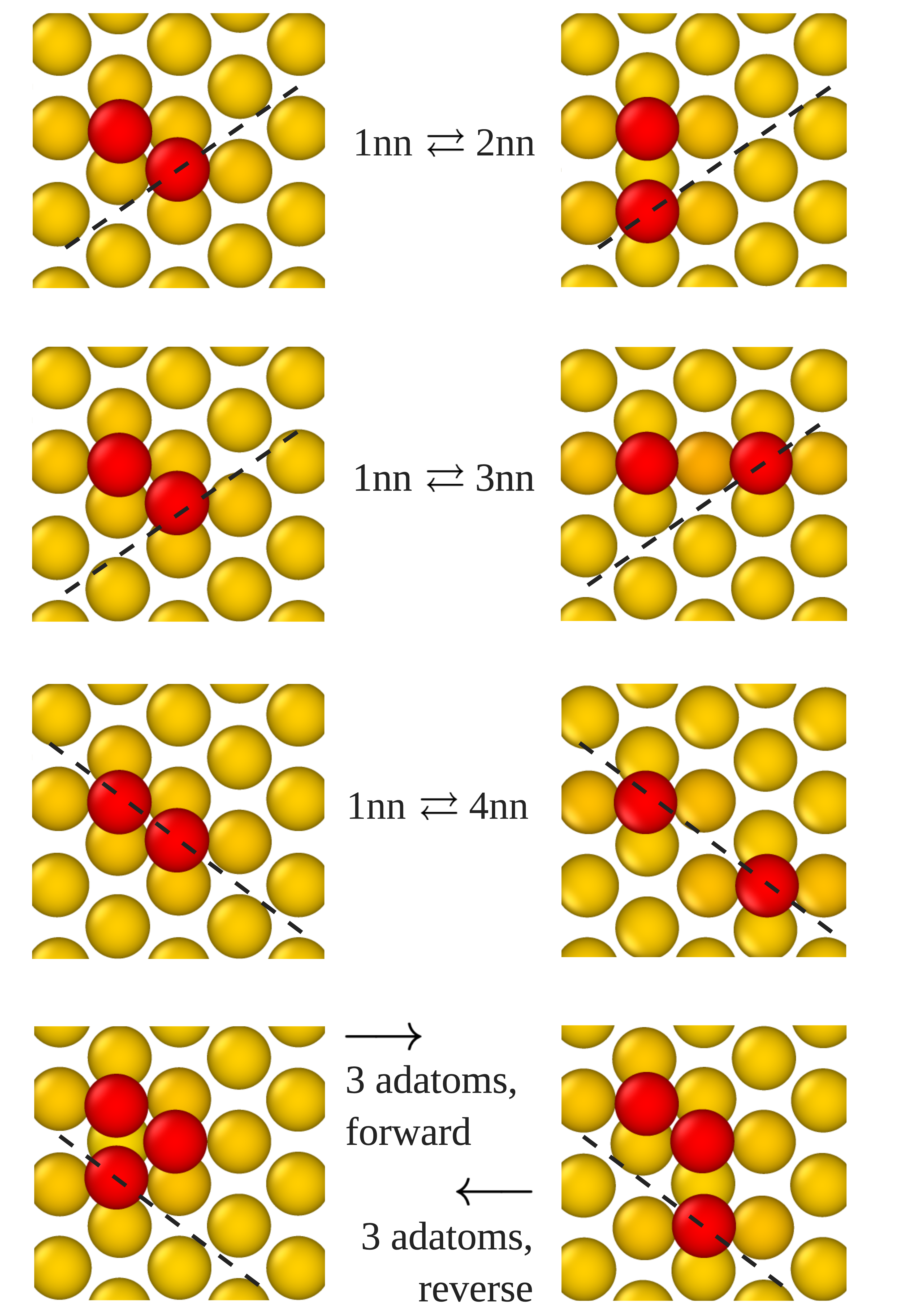}
\label{fig:jumps}
\end{figure}

To gain an understanding of how the LAE of the diffusing adatom affects its migration barriers in the presence of a field, we studied eight cases of  adatom jumps to the various first-nn lattice positions. Fig. \ref{fig:jumps} depicts the initial and final configurations of each jump in the forward (from left to right) and the reverse (from right to left) directions. Below we discuss the saddle point calculations for the jumps under study.

In \cite{kyritsakis2019atomistic}, we estimated a saddle point for the adatom hop-on 1nn jump to be in the middle of the bridge site; it was fixed in $x$ and $y$ directions while being allowed to relax along $z$ during DFT with and without an applied electric field. Since the hop-on 1nn jump is a symmetric process on a W \{110\} surface and the jumping adatom did not have any neighboring adatoms in our calculations, the middle point assumption for a saddle position was valid. In this work, all the jumps we studied are asymmetric between their initial and final states. Thus we used the Nudged Elastic Band (NEB) method coupled with DFT or a molecular dynamics (MD) simulations framework to estimate the position of  saddle points.

For the \textit{3 adatoms, forward} jump, we used the climbing image nudged elastic band (CI-NEB) method \cite{henkelman2000climbing,jonsson1998nudged} incorporated in the VASP package. 
As NEB with DFT relaxation is a computationally heavy approach for surface calculations, we used only one linearly interpolated image between relaxed initial and final structures. 
The spring constant used for the elastic band was 5 eV$\text{\AA}^{-2}$ and atomic relaxation was performed using damped molecular dynamics with a scaling constant (POTIM) of 0.05.

We first run CI-NEB DFT calculations without electric field and then for four values of the electric fields up to 1 GV/m for both the anode and the cathode cases (two values per electric field sign). These calculations were computational heavy. Thus,
we tested an approach with a fixed saddle point relaxation: we took a relaxed saddle point image, which was found with CI-NEB DFT under zero electric field, fixed the jumping adatom position in the $x$ and $y$ directions, allowed it to relax in the $z$ direction, allowed other ions to relax in all of the directions, and ran DFT simulations under the same values of the electric field. 
We obtained an average of 0.01 meV difference in the ground state energies of the saddle point system between the CI-NEB DFT relaxation and the fixed saddle point relaxation, which is an acceptable convergence. 
We then ran more electric field cases for up to $\pm$ 1.5 GV/m with the fixed saddle point relaxation.

As we found that CI-NEB DFT calculations are extremely computationally heavy for our systems under study, we had to find a less computationally demanding approach of estimating saddle point positions for the rest of the jumps in \FD{Fig. \ref{fig:jumps}}. 
We, therefore, used the Gaussian approximation (GAP) machine learning potential from \cite{byggmastar2019machine,byggmastar2020gaussian} to find the saddle points of $1nn\rightarrow2nn$, $1nn\rightarrow3nn$,$1nn\rightarrow4nn$ jumps without an electric field. 
The atomic positions that correspond to the saddle point states were found by performing CI-NEB calculations with the LAMMPS MD package \cite{plimpton1995fast}; a total of 12 linearly interpolated images between pre-relaxed initial and final positions was used in each case. 
Saddle point atomic positions were then used to perform DFT calculations under the electric field. We again used the approach with a fixed saddle point relaxation. 
As in our previous work, we assumed that the saddle point position does not change in the $x$ and $y$ directions when an electric field is applied. 
As our test for the \textit{3 adatoms} system showed, this assumption is valid within the approximation of $\pm$ 0.01 meV for the ground energy relaxation.

\subsection{Calculations of energy barriers}\label{subsec:methods_barriers}
Up to 14 values of an applied electric field in the range of $\pm$1.5 GV/m were used to construct $E$--$F$ curves for each saddle point of the jumps under the study and extract the systemic dipole moments and polarizabilities of the saddle points ($\mathcal{M_{\text{s}}}$ and $\mathcal{A_{\text{s}}}$). 
We assumed that saddle points remained the same for the reverse jumps $1nn\leftarrow2nn$, $1nn\leftarrow3nn$,$1nn\leftarrow4nn$, and \textit{3 adatoms, reverse}  as in the case of forward jumps \FD{(see Fig. \ref{fig:jumps})}.

To calculate the effective polarizability and dipole moment, the reference point systemic dipole moment and polarizability ($\mathcal{M_{\text{r}}}$ and $\mathcal{A_{\text{r}}}$) are required together with $\mathcal{M_{\text{s}}}$, $\mathcal{A_{\text{s}}}$, $\mathcal{M_{\text{l}}}$, $\mathcal{A_{\text{l}}}$ according to Eq. \eqref{eq:effective_characteristic}. 
In our previous work, we used a flat slab system as a reference for the adatom hop-on jump. 
In this work, the \textit{adatom} system was used as a reference for the jumps $1nn\rightleftarrows2nn$, $1nn\rightleftarrows3nn$,$1nn\rightleftarrows4nn$; the $1nn$ system (Fig. \ref{fig:systems}) was used for the three adatoms forward and reverse jumps.

The effective dipole moments and polarizabilities for the jumps under study along with zero field energy barriers are listed in Table \ref{table:effective}. The error estimates for dipole moments and polarizabilities were calculated according to the error propagation rule applied to the error estimates of the systemic polarization characteristics given in Table \ref{table:systemic}.

The values in Table \ref{table:effective} were used to calculate the energy barriers using Eq. \eqref{eq:barrier_effective} for both cases of the uniform ($\Delta F = 0$) and non-uniform ($\Delta F \ne 0$) applied electric field (see Fig. \ref{fig:barriers+jumps}).
The DFT values of $E_m$ were calculated using the equation from \cite{kyritsakis2019atomistic}:

\begin{equation} \label{eq:barriers_gradient_dft}
  E_m \approx \left(E_s \left( F_s \right) - E_r \left( F_s \right) \right) - \left( E_l \left( F_l \right) - E_r \left( F_l \right) \right) \textrm{,}
\end{equation}

where $E_s$ is the energy of the system with adatom at the saddle position under the applied field $F_s$, $E_l$ -- the energy of the system with adatom at the lattice position under the applied field $F_l$, $E_r$ -- the energy of the reference system where the jumping adatom is absent.

\section{Results}

Results of the DFT calculations of systemic dipole moments and polarizabilities of the various systems studied in this work are presented in Table \ref{table:polarizability}. 

The effective polarizabilities for the same systems are summarized in Table \ref{table:effective}. 

Below we use the values from Tables \ref{table:polarizability} and \ref{table:effective} to draw important conclusions about the approximation of the effective polarization characteristics with the partial ones, the effect of the LAE on $\mu$ and $\alpha$ of the adatom, and the atomic diffusion under electric field in general.

 \begin{table}[h!]
\centering
 \caption{The systemic dipole moment $\mathcal{M}_l$ and the systemic polarizability $\mathcal{A}_l$ of different systems depicted in Fig. \ref{fig:systems} (for \textit{3 adatoms, reverse} system, see Fig. \ref{fig:jumps}) with adatoms at lattice positions.}
 \makeatletter{\renewcommand*{\@makefnmark}{}
 \footnotetext{*Adatom results were originally published in \cite{kyritsakis2019atomistic}.}\makeatother}
\label{table:polarizability}
\begin{tabular*}{\columnwidth}{@{\extracolsep{\fill}} l l l l }
\hline 
System  & $\mathcal{M_{\text{l}}}$ [e$\text{\AA}$]  & $\mathcal{A_{\text{l}}}$ [e$\text{\AA}^2$/V]     \\
 \hline
adatom*                   & 0.305 $\pm$ $10^{-4}$   & 27.74  $\pm$ $10^{-3}$\\ 
1nn                   & 0.473 $\pm$ $10^{-4}$   & 28.011  $\pm$ $10^{-3}$\\ 
2nn                   & 0.482 $\pm$ $10^{-3}$   & 27.982  $\pm$ $10^{-2}$\\ 
3nn                   & 0.592 $\pm$ $10^{-3}$   & 27.942  $\pm$ $10^{-2}$\\ 
4nn                   & 0.587 $\pm$ $10^{-4}$   & 27.948  $\pm$ $10^{-2}$\\ 
3 adatoms,forward                   & 0.48 $\pm$ $10^{-4}$   & 28.30  $\pm$ $10^{-2}$\\ 
3 adatoms,reverse                   & 0.62 $\pm$ $10^{-4}$   & 28.238  $\pm$ $10^{-3}$\\ 
4 adatoms                   & 0.479 $\pm$ $10^{-4}$   & 28.30  $\pm$ $10^{-2}$\\ 
\hline
\end{tabular*}
\label{table:systemic}
\end{table}

 \begin{table*}[t!]
 \centering
 \scriptsize
 \caption{Effective atomic dipole moments and polarizabilities of the adatom in various LAE for the atomic jumps in Fig. \ref{fig:jumps} }
 \makeatletter{\renewcommand*{\@makefnmark}{}
 \footnotetext{$\mu_{s}$,  $\alpha_{s}$ are the effective atomic dipole moments and polarizabilities of the adatom at saddle points; $\mu_{l}$, $\alpha_{l}$ -- at the lattice positions; $\Delta \mu = \mu_{s} - \mu_{l}$, $\Delta \alpha = \alpha_{s} - \alpha_{l}$; $E_0$ are the energy barriers of the jumps without field. $\rightarrow$ indicates a forward jump, $\leftarrow$ indicates a reverse jump. Fitted values of zero field energies from energy vs electric field curve are used for calculating $E_0$ barriers. Partial dipole moments $m^{(a)}_x$ and polarizabilities $a^{(a)}_x$ (see Fig. \ref{fig:bader}) of the adatom were used in the construction of the ``partial'' row. Forward and reverse jumps are the same in the case of adatom and partial charges adatom systems. *Adatom results were originally published in \cite{kyritsakis2019atomistic}.}\makeatother}
\label{table:effective}

\begin{adjustbox}{width=\textwidth}

\begin{tabular}{*9l}

 \toprule[0.5pt]
 \vspace{5pt}
System  & $\mu_{s}$ [e$\text{\AA}$]  & $\alpha_{s}$ [e$\text{\AA}^2$/V]   & $\Delta \mu^{\rightarrow}$ [e$\text{\AA}$]  & $\Delta \alpha^{\rightarrow}$ [e$\text{\AA}^2$/V] & $\Delta \mu^{\leftarrow}$ [e$\text{\AA}$]  & $\Delta \alpha^{\leftarrow}$ [e$\text{\AA}^2$/V] & $E^{\rightarrow}_0$ [e$V$] & $E^{\leftarrow}_0$ [e$V$] \\
 \midrule[0.5pt]
 \\
  \vspace*{5pt}
adatom*                                      & 0.274 $\pm$ $10^{-3}$   & 0.261  $\pm$ $10^{-2}$   & -0.032  $\pm$ $10^{-3}$   & 0.031  $\pm$ $10^{-2}$    &  -  &  -  & 0.937 & -\\ 
\vspace*{5pt}
partial                                     & 0.269 $\pm$ $10^{-3}$   & 0.249  $\pm$ $10^{-2}$   & -0.068  $\pm$ $10^{-3}$   & -0.002  $\pm$ $10^{-2}$   &  -  &  -  & 0.937 & -\\ 
\vspace*{5pt}
1nn $\rightleftarrows$ 2nn                   & 0.157 $\pm$ $10^{-2}$   & 0.133  $\pm$ $10^{-1}$   & -0.010  $\pm$ $10^{-2}$   & -0.139  $\pm$ $10^{-1}$   & -0.019  $\pm$ $10^{-2}$    & -0.109  $\pm$ $10^{-1}$   & 1.644 & 1.033\\ 
\vspace*{5pt}
1nn $\rightleftarrows$ 3nn                   & 0.205 $\pm$ $10^{-3}$   & 0.291  $\pm$ $10^{-2}$   & 0.038  $\pm$ $10^{-3}$    & 0.019  $\pm$ $10^{-2}$    & -0.081  $\pm$ $10^{-3}$    & 0.088  $\pm$ $10^{-2}$    & 1.681 & 1.564\\ 
\vspace*{5pt}
1nn $\rightleftarrows$ 4nn                   & 0.271 $\pm$ $10^{-2}$   & 0.238  $\pm$ $10^{-1}$   & 0.104  $\pm$ $10^{-2}$    & -0.034  $\pm$ $10^{-1}$   & -0.011  $\pm$ $10^{-2}$    & 0.030  $\pm$ $10^{-1}$    & 1.435 & 1.178\\ 
\vspace*{5pt}
3 adatoms                                    & 0.085 $\pm$ $10^{-3}$   & 0.334  $\pm$ $10^{-2}$   & 0.079  $\pm$ $10^{-3}$    & 0.046  $\pm$ $10^{-2}$    & -0.064  $\pm$ $10^{-3}$    & 0.107  $\pm$ $10^{-2}$    & 0.889 & 0.417\\
\\
\bottomrule[0.5pt]

\end{tabular}
\end{adjustbox}
\end{table*}

\subsection{Partial polarization characteristics}\label{subsec:results_partial}

To evaluate whether the partial polarization characteristics can be used for describing the diffusion under the applied electric field, i.e. if the effective dipole moment and polarizability of the adatom could be approximated by the corresponding partial values, we first consider the case of a uniformly applied electric field ($\Delta F =0$ in Eq. \eqref{eq:barrier_effective}).

\begin{figure}[!t]
\caption{Energy barriers of the adatom calculated with the effective dipole moment and polarizability and the partial dipole moment and polarizability under a uniform electric field. Partial polarization characteristics calculations are explained in Fig. \ref{fig:bader}. Gray shadowed regions correspond to the error margin calculated by the error propagation
rule applied on the uncertainties of the parameters given in Table \ref{table:effective}. The uncertainty for $E_0$ was set to $0.1$ meV.}
 \centering
\includegraphics[width=0.48\textwidth]{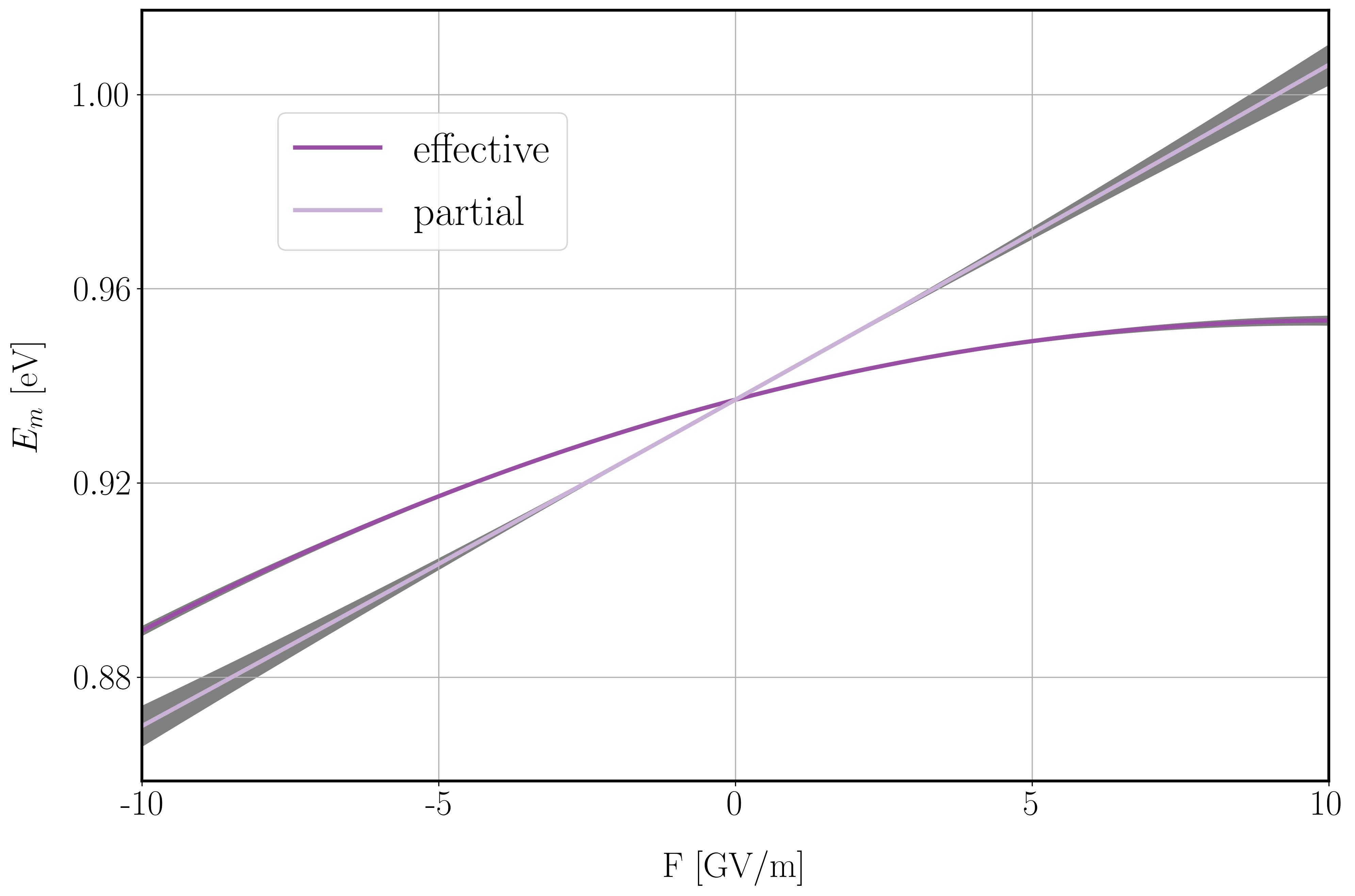}
\label{fig:adatom_vs_isolated_homog}
\end{figure}

From Table \ref{table:effective}, we can see that whereas $\Delta \mu$ of the adatom system is comparable with the partial one, the polarizabilities $\Delta \alpha$ differ significantly. In fact, the partial $\Delta a^{(a)}$ is so small that it could be approximated as zero. This makes the parabolic dependency $E_m(F)$ impossible in this approach, which is crucial at the high values of an applied electric field. This point is illustrated in Fig. \ref{fig:adatom_vs_isolated_homog}, where we can see the linear trend and the enlarged shadowed uncertainty regions at fields $> 5$ $GV/m$ in magnitude for the partial curve.

\begin{figure}[t]
\caption{Energy barriers of the adatom calculated with the effective polarization characteristics and with the partial ones under a non-uniform electric field. Text boxes next to the graphs specify the applied electric field values. Negative values correspond to a cathode, positive --  to an anode}
\label{fig:adatom_vs_isolated_heterog}

 \centering
\includegraphics[width=0.48\textwidth]{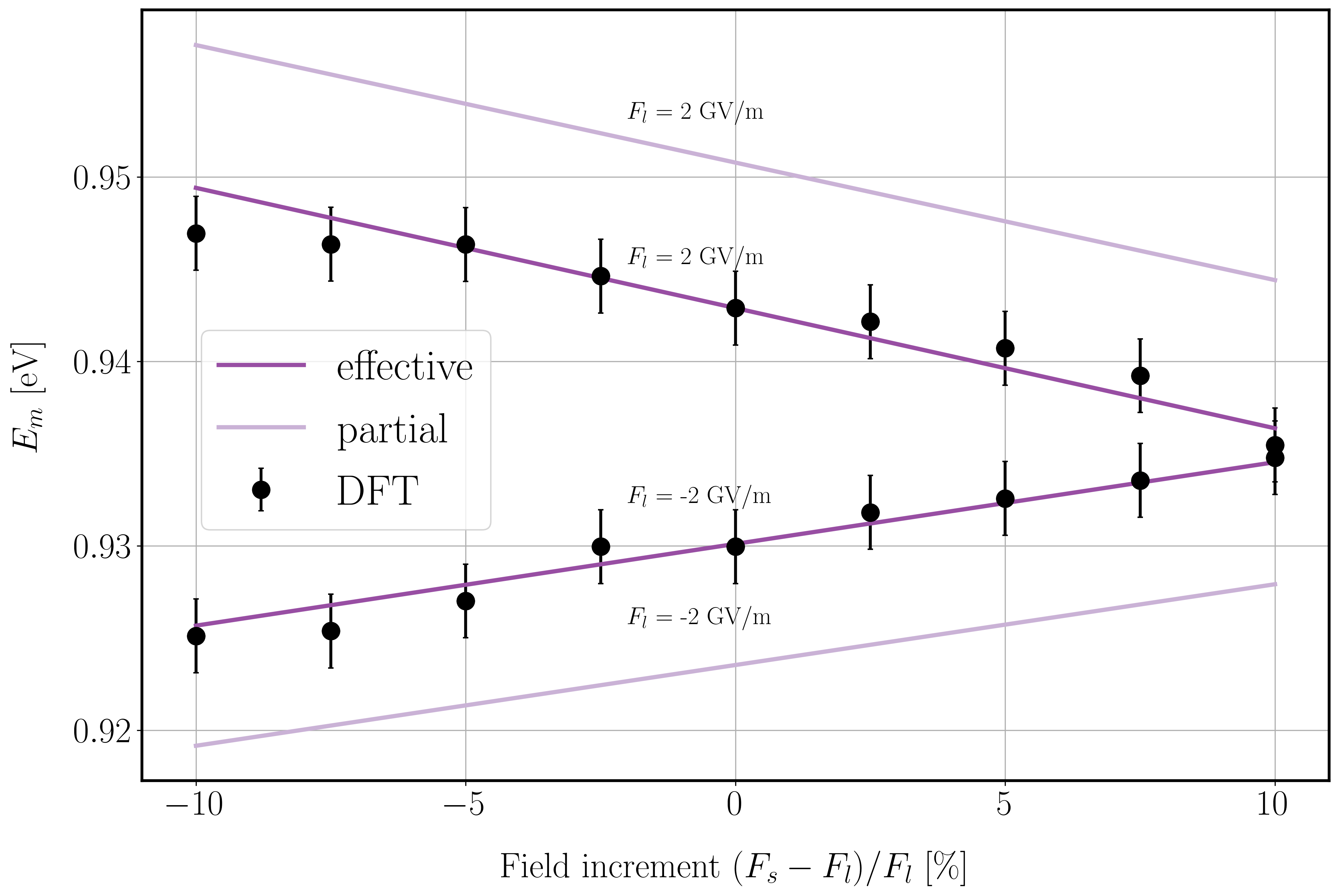}
\end{figure}

However, if we now consider the diffusion under a field gradient, where the effective polarization characteristics of the adatom at the saddle point ($\mu_s$ and $\alpha_s$) start to play a crucial role, especially in the case of large field gradients $\Delta F$, we can notice that the effective values for the saddle point position of the adatom are comparable with the partial ones. 
This point is reflected by the same trends of $E_m$ dependency on the field gradient $\Delta F$ in Fig. \ref{fig:adatom_vs_isolated_heterog}. 
The comparability of the effective and partial polarization characteristics of the adatom at the saddle point can be explained intuitively: at the saddle point, the adatom is located on top of the bridge site, which results in less charge transfer from surface atoms to the adatom than in the case of a lattice position, where the adatom sits in the hollow site between four surface atoms, and thus it is affected more significantly by their presence.  

Moreover, we can see in Fig. \ref{fig:adatom_vs_isolated_homog} that the diffusion is significantly underestimated ($E_m$ is much higher) while using the partial polarization characteristics on the anode side ($F > 0$) and overestimated ($E_m$ is lower) on the cathode side ($F < 0$). 
However, in Fig. \ref{fig:adatom_vs_isolated_heterog} we see that the diffusion bias, which is reflected by the slope of the curves, is predicted to be similar in both approaches. 
\subsection{Effective polarization characteristics of adatoms in various LAE}\label{subsec:results_effective}
\begin{figure*}
  \caption{Effective dipole moment $\mu_l$ and polarizability $\alpha_l$ of adatoms at the lattice sites in various LAE. For all cases, the reference system had one adatom (which polarization characteristics are presented) less. The corresponding configurations of the reference systems are given in the parentheses. Adatoms for which $\mu_l$ and $\alpha_l$ are calculated are highlighted with dashed squares. In the case of two adatoms systems, both adatoms have the same $\mu_l$ and $\alpha_l$ }
  \centering
    \includegraphics[width=\textwidth]{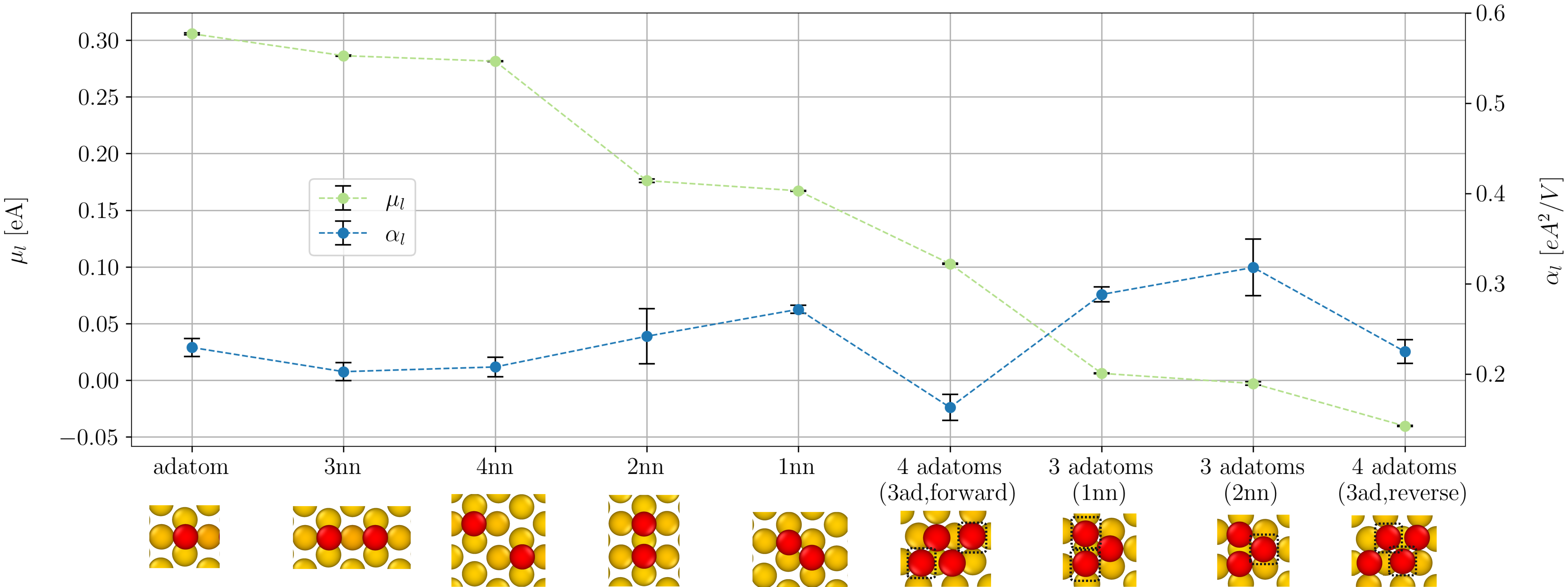}
\label{fig:mu_alpha_lae}
\end{figure*}

Figure \ref{fig:mu_alpha_lae} displays the effective dipole moments and polarizabilities of adatoms at lattice positions in different local atomic environments. 
The figure is sorted by the decreasing dipole moment. 
It starts from the one adatom system, where the dipole moment is the highest due to the strongest charge redistribution around the adatom and ends with the islands of three and four adatoms where the charge redistribution on the jumping adatom is affected the least and thus the effective dipole moment is close to zero. We show the polarization characteristics separately only for those adatoms, which are outlined with the dashed square in the atomistic image of the corresponding configuration below. 
The negative value of the effective dipole moments of ``inner'' adatoms (last image) in the \textit{4 adatoms} system means that the systemic dipole moment is lower than the dipole moment of the reference system (initial position of the \textit{3 adatoms, reverse} jump in this case).

By looking at the effective dipole moments and effective polarizabilities of the adatoms in the \textit{adatom}, \textit{3nn}, \textit{4nn}, \textit{2nn}, \textit{1nn} systems, we can deduce that $\mu$ and $\alpha$ are significantly affected by other adatoms in the LAE if the latter are located within the 2nn distance. If the neighboring adatoms are located further than the 2nn distance, the effect of their charge on the jumping adatom might be neglected.

Another interesting point that can be seen in Fig. \ref{fig:mu_alpha_lae} is the similarity between $\mu_l$ and $\alpha_l$ of all the adatoms in the three adatoms island. 
This is however not the case for the four adatoms island, where the ``inner'' and ``outer'' jumping adatoms have very different effective polarizabilities and dipole moments. 

In general, we can conclude that the polarization characteristics of adatoms are significantly affected by their LAE, which is expected, as we have already established the importance of including the vicinity of the adatom in the calculations of effective dipole moment and polarizability. 

\subsection{Atomic diffusion under electric fields}\label{subsec:results_diffusion}
\begin{figure*}
  \caption{ Right graph: Difference in the energy barrier introduced by the uniformly applied electric field for $1nn \rightleftarrows 3nn$, \textit{3 adatoms, forward} and \textit{3 adatoms, reverse} jumps. $F>0$ region corresponds to an anode, $F<0$ to a cathode. Left graph: energy barriers under the non-uniformly applied electric field for the same jumps. The x-axis of the left graph is the relative field increment $(F_s - F_l)/F_l$ for the  $\pm$ 2 GV/m applied field. Positive increment indicates the increasing field toward the saddle point (for both cases of the field sign). When the increment is negative, the magnitude of the field is higher at the lattice site than at the saddle point. Applied fields are shown next to the corresponding lines. Schematics of jumps are shown in the middle with the legend explaining colors and line types of the corresponding graphs. Both graphs also include the adatom jump, i.e. jump with no neighbors present. Shadowed regions around the curves correspond to the error margin calculated by the error propagation rule.}
  \centering
    \includegraphics[width=\textwidth]{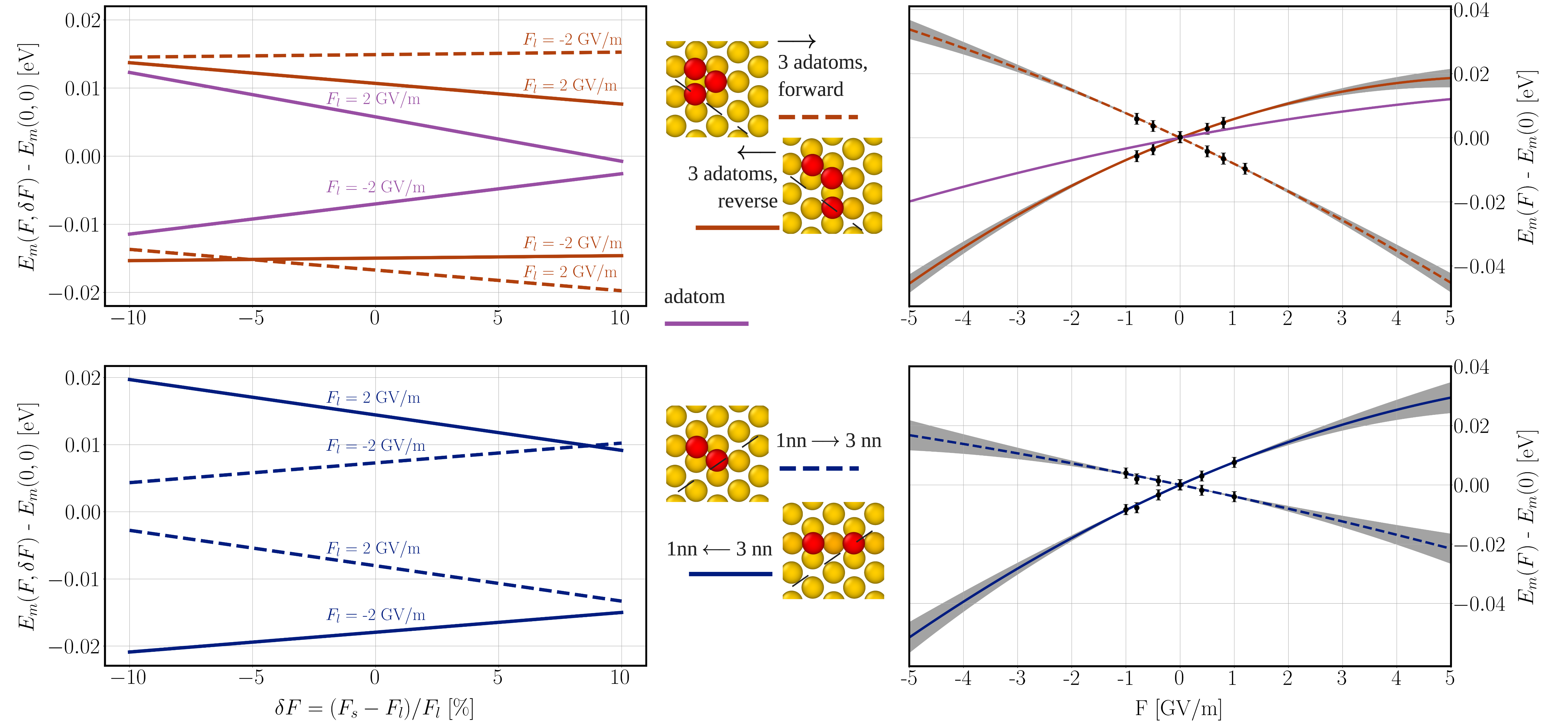}
\label{fig:barriers+jumps}
\end{figure*}
We can  conclude that diffusion is modified differently in different cases of LAE as both the effective saddle point parameters and the differences between lattice and saddle parameters differ not only by magnitude but also by sign, meaning that the diffusion trends will be opposite for certain combinations of the applied field and field gradient (see Table \ref{table:effective}). Below we only discuss the energy barriers of the $1nn\rightleftarrows3nn$, \textit{3 adatoms, forward} and \textit{3 adatoms, reverse} jumps due to the high uncertainties in the other cases. 

Figure \ref{fig:barriers+jumps} shows the diffusion trends under relatively small fields (up to 10 GV/m). 
We also included a single adatom jump with no neighboring adatoms in LAE (marked as \textit{adatom}) for comparison. 

Under uniform applied electric field (right graph in Fig. \ref{fig:barriers+jumps}) on the anode side ($F > 0 $), \textit{3 adatoms, forward} and $1nn\rightarrow3nn$ jumps become promoted and their reverse jumps become suppressed. The adatoms tend to separate. On the cathode side ($F < 0 $), the trend is opposite: the adatoms tend to form a cluster. 
Both trends become prominent with the field gradient on the anode side, since the barriers are lower towards higher electric fields ($F_l < F_s $) for all five jumps shown schematically on the left graph of Fig. \ref{fig:barriers+jumps}. All the jumps also have the same trends on the cathode side, but in the opposite direction than on the anode: jumps are promoted towards lower fields ($F_l > F_s $). 

\section{Discussion}
\subsection{Effective and partial polarization characteristics}
In \cite{kyritsakis2019atomistic}, we showed that to accurately describe the atomic dynamics of a surface under a high electric field, the systemic dipole characteristics are necessary. 
The disadvantage of that description is that it is based on non-atomic quantities, making the expressions for the calculation of forces and barriers relatively cumbersome and counter-intuitive. 
In contrast, the atomic partial polarization characteristics describe the surface dynamics under electric fields in atomic terms, i.e. using quantities that describe each atom, giving a simpler and more elegant description. 
However, it is not accurate, as shown in Sec. \ref{subsec:results_partial}.
The concept of the effective polarization characteristics introduced in Sec.  \ref{subsec:theory_effective} combines the advantages of both pictures, i.e it uses atom-bound quantities and provides high accuracy. 

\subsection{Dependence on the LAE and its implications}
In \cite{jansson2020growth}, we simulated the atomic surface diffusion on a W surface under electric fields using KMC. Since we only knew the values of $\mu_s$, $\alpha_s$, $\Delta\mu$ and $\Delta\alpha$ (corresponding to $\mathcal{M}_{sr}$, $\mathcal{A}_{sr}$, $\mathcal{M}_{sl}$, $\mathcal{A}_{sl}$, respectively, in \cite{jansson2020growth}) for one single process, the adatom jump on a \{110\} surface, we approximated all jumps with these values. In practice, we were ignoring any dependency of these parameters on the LAE. Only the zero-field barrier and the applied field for every jump would vary in the model. We were able to predict that nanotips would start to grow from the surface if the fields were high enough. The growth was observed even if the parameters were varied by $\pm$20 \%. If the LAE was taken into account, which would require using DFT to calculate the $\mu_s$, $\alpha_s$, $\Delta\mu$ and $\Delta\alpha$ parameters for about 500--1000 surface processes \cite{jansson2020tungsten}, how much precision would such a model gain? From Table \ref{table:effective}, we can see that the parameters may vary much more than 20 \% compared to the adatom values (the same values as were used in \cite{jansson2020growth}). Even though we have only studied four new processes, the parameters vary significantly compared to the adatom case and may even change the signs. This would indicate that calculating the effective atomic polarization characteristics parameters for all 500--1000 surface processes would significantly increase the precision of the model in \cite{jansson2020growth}, although this is beyond the scope of this paper.

Finally, we should note that although the results in this work show that $\mu$ and $\alpha$ have a significant dependence on LAE, we found in Sec.  \ref{subsec:results_effective} that only neighboring adatoms within 2nn distance affect $\mu_l$ and $\alpha_l$. This observation allows for a significant reduction of the number of LAE combinations needed for the parameterization of a KMC model. We note, nevertheless, that the presence of the neighbors further than 2nn might still play a role if these are connected to the atoms in the immediate LAE of the jumping atom.

From the limited number of cases we studied here, it was not possible to deduce a general trend for the LAE dependence. Thus, it has to be explored further in order to investigate its implications on larger-scale simulations such as Molecular Dynamics (MD) and Kinetic Monte Carlo (KMC).

\subsection{Diffusion trends}
The results in Sec. \ref{subsec:results_diffusion} indicate that for the studied range of electric fields, a cathode field promotes adatom island creation since the barriers are lowered for adatoms jumping towards each other. On the contrary, the anode field slightly lowers the barriers for adatoms to separate from each other. Note, however, that the separation barrier is significantly higher than the reverse barrier regardless of the field. More cases of adatom islands need to be studied to confirm the electric field effect on adatom island formation.

Finally, we note that the uncertainties for some jumps that were calculated with a fixed saddle point approach are rather large, which indicates that either the saddle point positions were not estimated precisely with the GAP potential, or that the saddle point state is modified significantly by the field so that the fixed saddle point approach becomes a crude approximation. We also note that due to the parabolic dependence of Eq. \eqref{eq:barrier_effective}, starting from a certain value of the applied electric field on the cathode side (e.g. -10 GV/m for a single adatom jump), the trend becomes reversed. However, in this work, we do not discuss high electric fields due to the aforementioned uncertainties becoming significant starting from fields up to 5 GV/m, as indicated by the shadowed error regions on the right graph of Fig. \ref{fig:barriers+jumps}.

\section{Conclusions}
We rigorously defined the concept of the effective atomic polarization characteristics — permanent dipole moment and polarizability. These quantities describe the atomic dynamics of metal surfaces under a high electric field.
The introduction of these quantities results in a concise and accurate description of the surface dynamics, as it is shown to be mathematically equivalent to our previous theory of the systemic polarization characteristics while maintaining the simplicity and compactness of atomic parameters, i.e. parameters that describe an atom at a given local environment, rather than a whole system. 

We investigated the applicability of approximating the effective polarization characteristics by the partial atomic ones, which were previously suggested within a simplified approach to address the problem of biased surface diffusion on metal surfaces under electric fields. We show that although the effective and partial atomic dipole characteristics have similar values, using the partial values induces significant errors in the calculation of the migration barriers.

Finally, we showed that the local atomic environment of an adatom affects its effective polarization characteristics and hence its migration energy barriers, by studying a limited number of local atomic environments of the adatom. For a future KMC model of the atomic surface evolution under electric fields, this would indicate that a significant improvement in precision would be gained if the LAE is taken into account for all surface atom jump processes when calculating their effective atomic polarization characteristics.

\newpage
\section*{Acknowledgements}
 E.\;Baibuz was supported by a CERN K-contract and the doctoral program MATRENA of the University of Helsinki. A. Kyritsakis by the same K-contract (47207461) and European Union’s Horizon 2020 program, under Grant No. 856705 (ERA Chair ‘MATTER’). F.\;Djurabekova acknowledges gratefully the financial support of Academy of Finland (Grant No. 269696).
 
\bibliographystyle{elsarticle-num}
\bibliography{dft_bib}
\newpage

\appendix
\section{Symbols and notations of various polarization characteristics}

We have introduced three sets of polarization characteristics in this paper: \textit{systemic}, \textit{partial}, and \textit{effective}. For reader's convenience, we compile all the relevant terms for different kinds of dipole moments and polarizabilities and their  notations in Table \ref{table:symbols} 

 \begin{table}[h!]
\centering
 \caption{Notations of various polarization characteristics}
\label{table:symbols}
\begin{tabular*}{\columnwidth}{@{\extracolsep{\fill}} l l }
 \toprule
Symbol  & Description     \\
 \midrule
$\mathcal{M}$                        & systemic dipole moment \\
$\mathcal{A}$                        & systemic polarizability \\ 
$\mathcal{P}$                        & systemic dipole moment under the electric field \\
$\mathcal{M}_s$, $\mathcal{M}_l$     & systemic dipole moment of a slab with \\
                                     & the adatom at the saddle and lattice positions \\
$\mathcal{M}_r$                      & systemic dipole moment of a slab \\
                                     & without the jumping adatom \\
$\mathcal{M}_{sl}$                   & $\mathcal{M}_s$ - $\mathcal{M}_l$\\
$\mathcal{M}_{sr}$                   & $\mathcal{M}_s$ - $\mathcal{M}_r$\\
$\mathcal{A}_s$,$\mathcal{A}_l$      & systemic polarizability of a slab with \\
                                     & the adatom at the saddle and lattice positions \\ 
$\mathcal{A}_r$                      & systemic polarizability of a slab \\
                                     & without the jumping adatom \\
$\mathcal{A}_{sl}$                   & $\mathcal{A}_s$ - $\mathcal{A}_l$ \\
$\mathcal{A}_{sr}$                   & $\mathcal{A}_s$ - $\mathcal{A}_r$ \\

$\mu_{x}$                            & effective dipole moment of the adatom \\
                                     & at a random position $x$ \\ 
$\mu_{s}$,$\mu_{l}$                  & effective dipole moment of the adatom \\
                                     & at the saddle and lattice positions \\ 
$\alpha_{x}$                         & effective polarizability of the adatom \\
                                     & at a random position $x$ \\ 
$\alpha_{s}$,$\alpha_{l}$            & effective polarizability of the adatom \\
                                     & at the saddle and lattice positions \\ 
$\Delta \mu$                         & $\mu_{s}$ - $\mu_{l}$ \\ 
$\Delta \alpha$                      & $\alpha_{s}$ - $\alpha_{l}$\\ 

$m^{(a)}$                            & partial dipole moment of the adatom  \\  
$a^{(a)}$                            & partial polarizability of the adatom \\
$p^{(a)}$                            & partial dipole moment of the adatom \\
                                     & under the electric field\\  
\bottomrule
\end{tabular*}
\end{table}

 \end{document}